\documentclass[epsf]{mn2e}
\def\mmm{(m-M)$_0$}
\def\ebv{$E(B-V)$~}

\def\gsim{\;\lower.6ex\hbox{$\sim$}\kern-7.75pt\raise.65ex\hbox{$>$}\;}
\def\lsim{\;\lower.6ex\hbox{$\sim$}\kern-7.75pt\raise.65ex\hbox{$<$}\;}

\title[Cr 110]{Intermediate age open clusters: Collinder 110}

\author[Bragaglia \& Tosi]{Angela Bragaglia, Monica Tosi  \\
INAF--Osservatorio Astronomico di Bologna, Via Ranzani 1, I-40127 Bologna,
      Italy,
      e-mail angela@bo.astro.it, tosi@bo.astro.it \\
}

\date{}

\begin{document}
\maketitle

\begin{abstract}
We present CCD $BV$ photometry of the intermediate age open cluster Collinder
110, a nearby, scarcely populated, and poorly  studied system.
There is no literature information on the metallicity, so we tested several
possibilities, and found a slight evidence of sub-solar abundances.
Using the synthetic Colour - Magnitude Diagrams technique we estimate the
following parameters: 
age between 1.1 and 1.5 Gyr, reddening 0.38$\leq$ \ebv $\leq$0.45, distance
modulus \mmm ~between 11.8 and 11.9 if the cluster metallicity is solar, or
age between 1.2 and 1.7 Gyr, reddening 0.52$\leq$ \ebv $\leq$0.57, distance
modulus \mmm ~between 11.45 and 11.7 if the cluster metallicity is sub--solar.

\end{abstract}


\section{Introduction}

Open clusters are commonly believed to be excellent tracers of the Galactic
disc properties. They can be observed in the whole disc, their parameters (e.g.
distance, age, and metallicity) can be determined with an accuracy unreachable
for other disc objects except the nearest ones,  their ages cover the
whole interval from a few million years to about 10 Gyr, so they can be used to
study both the present day disc structure and its temporal evolution
(Janes \& Phelps 1994, Friel 1995, Tosi 2000, Bragaglia et al. 2002).
In particular, intermediate age and old open clusters (i.e., with ages older
than the Hyades) offer the best opportunity to trace the whole kinematical and
chemical history of our disc, once granted the existence of populous and
representative  samples, accurately and homogeneously analysed (for recent works
see e.g., Twarog, Ashman, \& Anthony-Twarog 1997; Carraro, Ng, \& Portinari
1998).

We are presently trying to build such a sample and to this end
we have collected data over several years. We have already analysed in a
homogeneous way eight clusters (see Di Fabrizio et al. 2001, and references
therein). This paper is part of our general project, and is devoted to Collinder
110 (Cr 110, C 0635+020), a poorly populated, intermediate age open cluster
located at RA(2000) = 06:38:35, DEC(2000) = +02:02:27, or l = 209.66,  b =
--1.99.

Cr 110 was previously studied by Dawson and Ianna (1998, hereafter DI98), who
presented  photographic and photoelectric measurements, and also reported
results of an older paper by Tsarevskii \& Abakumov (1971). They estimated
\ebv=0.50 $\pm$ 0.03, distance 1950 $\pm$ 300 pc, and age of 1.4 $\pm$ 0.3 Gyr
assuming solar abundance. Dutra and Bica (2000) cite a larger reddening 
(\ebv=1.12) based on the Schlegel, Finkbeiner \& Davis (1998) maps, which
however should be used with great caution at such low galactic latitudes (see
e.g., Appendix C of Schlegel et al., where they suggest not to trust their
predicted reddenings  for $|b| < 5^\circ$).

We describe our data in Section 2, and the resulting colour - magnitude
diagrams in Section 3. Section 4 is devoted to the derivation of the cluster 
parameters, while summary and conclusions are presented in Section 5.

\begin{figure}
\vspace{14cm}
\includegraphics{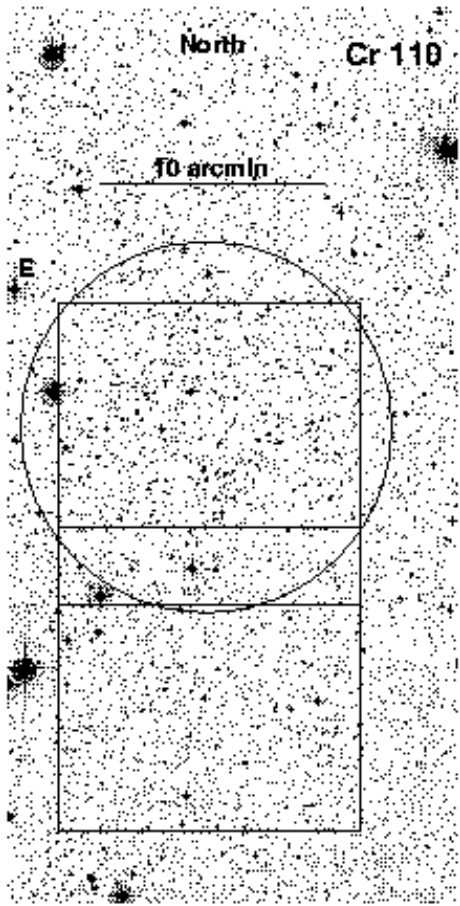}
\caption{Map of our field (the central square) of 13.5 arcmin$^2$;
also shown is a second field used for field stars comparison, and the zone
observed by Dawson \& Ianna (1998, fig. 1, the circle).}
\label{fig-map}
\end{figure}

\section{Observations and data reduction}

Cr 110 was observed with DFOSC (Danish Faint Object Spectrograph and Camera) at
the 1.54m Danish telescope located in La Silla, Chile, on UT January 5 and 6,
1999. DFOSC was equipped with CCD \#C1W7 (Loral/Lesser, 2k $\times$ 2k pixel,
0.4\arcsec/pixel, field of view of 13.7 arcmin$^2$), and the Johnson  standard
system filters.  
The two nights appeared photometric; seeing values varied between about 
1.5\arcsec and 1.8\arcsec. We obtained 6 exposures in the $B$ filter (10, 20, 
60, 120, 360, 600 s)  and 5 in the $V$ filter
(5, 20, 60, 180, 300 s) on a field centered on the cluster, and 2 exposures in
$B$ and $V$ on a second, partly overlapping field (20 and 240 s, 10 and 120 s in
the two bands respectively) we observed to  account for field stars
contamination.  Fig.~\ref{fig-map} shows our fields, and the DI98 one.

\begin{figure*}
\vspace{14cm}
\includegraphics{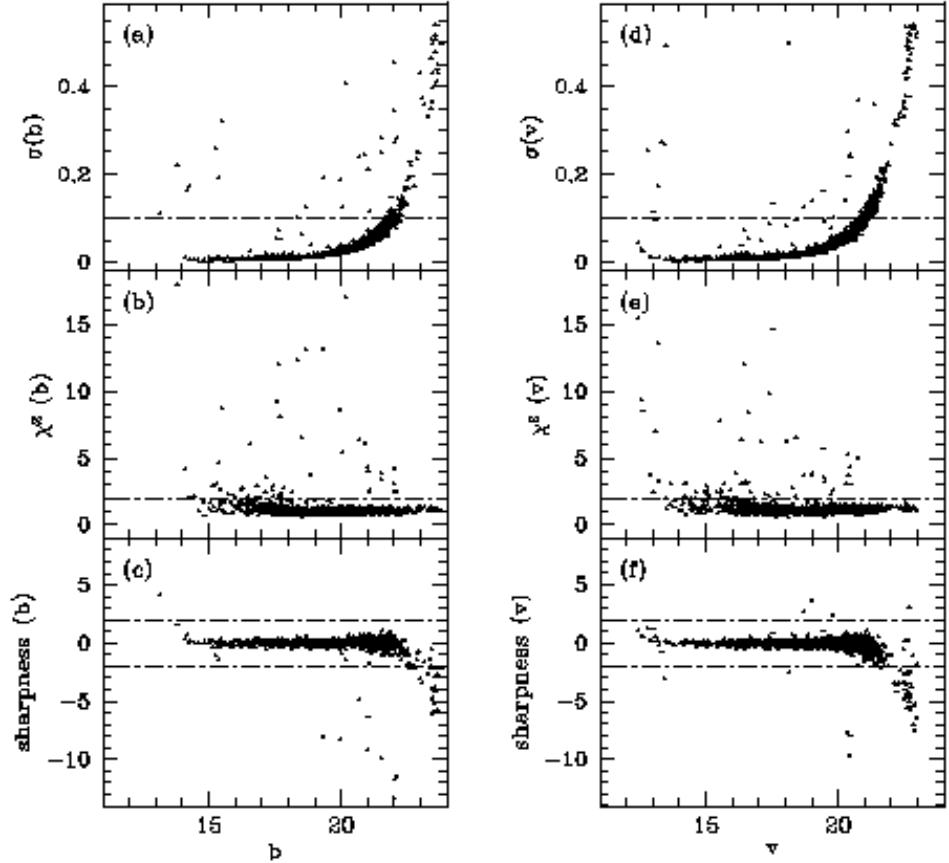}
\caption{Distribution of errors (the DAOPHOT $\sigma$), $\chi^2$, and
$sharpness$ for the B and V reference frames. Note that we plot the
instrumental $b$ and $v$ magnitudes.}
\label{fig-sele}
\end{figure*}

All frames were trimmed and corrected for bias and flat fields in the 
standard way, using IRAF\footnote{
IRAF is distributed by the NOAO, which are operated by AURA, under contract
with NSF} tasks. We then used 
DAOPHOT--II, also in IRAF environment, to find and measure
stars (Stetson 1987, Davis 1994).
All frames were searched independently, using the appropriate value for the
FWHM of the stellar profile and a threshold of 4 $\sigma$ over the local
sky value. 
About 30 well isolated, bright stars distributed all over the frame were used
in each image to define the best analytical PSF model (we used a gaussian with
spatial variations), which was then applied to all detected objects.
The resulting magnitude file was selected in magnitude, to avoid saturated
stars, in error (only stars with $\sigma \le$ 0.1 were retained, and almost
all rejections were in the lower magnitude bins),
in goodness-of-fit estimator ($\chi^2 \le $ 2),  and in shape - defining 
parameter ($-1 \le sharpness \le$ 1), to avoid cosmic rays and
false identifications of extended objects. Fig.~\ref{fig-sele} shows
the distribution in error, $\chi^2$, and $sharpness$ for the reference $B$ and
$V$ frames, with lines indicating our selection criteria.

We computed a correction to the PSF derived magnitudes to be on the
same system as the photometric standard stars:
aperture photometry was performed on a few isolated stars (the same used to
define the PSF) in the reference image for each filter.  
The corrections (in the sense aperture minus PSF)
were found to be --0.187 in $B$, and --0.223 mag in $V$.      

All output catalogues referring to the central field were aligned to the one
derived from  the reference B image, assumed as master frame for the
coordinate system, using dedicated programs developed at the Bologna
Observatory by P. Montegriffo. We then "forced" the  output catalogues in both
filters  to the reference ones applying linear transformations (almost  zero
point shifts) to the instrumental magnitude. 
The final magnitudes  in each band are the result of the
average of all measures for each star. 

Equatorial coordinates for all stars in the final catalogue were
computed using another dedicated program developed by P. Montegriffo, which
cross correlates positions (in pixel) of our stars with positions (in
RA and DEC) of objects in the second
edition of the Guide Star Catalogue,\footnote{
 The Guide Star Catalogue-II is a joint project of the Space Telescope Science
 Institute and the Osservatorio Astronomico di Torino}
and computes suitable transformations. 
The r.m.s. of the transformation was of about 0.1\arcsec in both coordinates.

The external field was also aligned in magnitude to the $B$ and $V$ reference
central fields, using stars in common, and formed a separate catalogue.

\subsection{Photometric calibration}

The conversion of the final, averaged catalogue from instrumental magnitudes 
to the Johnson standard system was obtained using the standard area PG0231+051
(Landolt 1992) observed four times just before and after the cluster on 
the first night.
The standards span a fairly wide range in colour (--0.329 $\leq B-V \leq$
1.448), adequate for the cluster stars. Aperture photometry was used to measure
their magnitudes.

Extinction coefficients were derived from the standard stars, since they were
observed at different airmasses: $\kappa_V=0.1354$ and $\kappa_B=0.2287$
(Clementini,  private communication).
No second order extinction coefficients were estimated, but they are 
usually close to zero, and can safely be ignored if one does not aim to
milli-mag accuracy (intrinsically impossible with the instrument
we used, which is not designed for very high precision photometry). 
Moreover our data were not acquired at large airmasses
so there is no strong need of a color-dependent extinction correction, which 
accounts for differential absorption at the blue and red ends of the filter 
passbands.  
The extinction coeeficients used in this work compare well to the average
values for the site ($\kappa_V \simeq 0.12, ~\kappa_B \simeq 0.22$)  and to the
ones  derived for another night of the same run (UT January 7 1999) using 
different objects i.e., stars in the Large Magellanic Cloud observed the  whole
night at very different airmasses ($\kappa_V = 0.114, ~\kappa_B = 0.252$,
Clementini, private communication). 

The instrumental magnitudes of the standard stars were corrected for 
extinction and to an exposure time of 1 second, and from comparison to the
tabulated value we derived the following calibration:

$$ B = b + 0.0999 \times (b-v) - 1.0633 ~~~~(r.m.s.=0.0114) $$
$$ V = v + 0.0063 \times (b-v) - 0.6075 ~~~~(r.m.s.=0.0163). $$

\noindent
These equations, 
where $B$ and $V$ are the Johnson magnitudes, and $b$ and $v$ are the
instrumental magnitudes, corrected for aperture, exposure time and extinction,
were then used to build our final photometric catalogue.

DI98 presented also photoelectric measurements, and we compared the $B$ and
$V$ values for the about 30 stars in common. We noticed a small trend, i.e., our
magnitudes are brighter than the photoelectric ones at the bright end, and
fainter at the faint end, but with a large scatter. We have derived a correction
and applied it to our catalogue; however, when comparing our corrected 
Colour - Magnitude diagram with the photographic one by DI98, calibrated using
the photoelectric measures, not only we did not find any
clear improvement, but a worsening was possible. We then decided not to apply 
any correction. 

\begin{table}
\begin{center}
\caption{Completeness ratios in the two bands.}
\vspace{5mm}
\begin{tabular}{ccc}
\hline\hline
   mag  &compl B &  compl V \\
\hline
$<$15.5 & 1.00 & 1.00 \\
  15.50 & 0.99 & 0.99 \\
  15.75 & 0.97 & 0.95 \\
  16.00 & 0.98 & 0.97 \\
  16.25 & 0.96 & 0.95 \\
  16.50 & 0.96 & 0.95 \\
  16.75 & 0.96 & 0.95 \\
  17.00 & 0.96 & 0.94 \\
  17.25 & 0.96 & 0.94 \\
  17.50 & 0.93 & 0.93 \\
  17.75 & 0.94 & 0.92 \\
  18.00 & 0.96 & 0.94 \\
  18.25 & 0.95 & 0.92 \\
  18.50 & 0.93 & 0.93 \\
  18.75 & 0.93 & 0.92 \\
  19.00 & 0.94 & 0.90 \\
  19.25 & 0.93 & 0.93 \\
  19.50 & 0.94 & 0.93 \\
  19.75 & 0.94 & 0.90 \\
  20.00 & 0.90 & 0.89 \\
  20.25 & 0.92 & 0.92 \\
  20.50 & 0.92 & 0.84 \\
  20.75 & 0.89 & 0.83 \\
  21.00 & 0.87 & 0.78 \\
  21.25 & 0.82 & 0.51 \\
  21.50 & 0.69 & 0.08 \\
  21.75 & 0.35 & 0.00 \\
  22.00 & 0.04 &  \\
  22.25 & 0.00 &  \\
\hline
\end{tabular}
\end{center}
\label{tab-compl}
\end{table}

\subsection{Completeness analysis}

We tested the completeness of our stellar detections on the deepest $B$ and 
$V$ images, adding artificial stars to each frame and exactly repeating the
procedure of extraction of objects and PSF fitting used for the original frame.
The stars were added at random positions 
and selected in magnitude according to the observed
luminosity  function. 
We added only 960 objects at a time, and the process was repeated as many times
as to reach a total of about 50,000 artificial stars. In fact, in order not to
significantly alter the crowding conditions we added each object in  a $60
\times 60$ pixel box, excluding the image borders. This way we   approximate
the condition of adding a single star each time, i.e. of a repeated,
independent experiment, since the artificial objects added in a single run do
not interfere with each other.  
To the output catalogue of the added stars we applied the same selection 
criteria in magnitude, error, $\chi^2$  and sharpness as done for the science
frames. The completeness degree of our photometry at each magnitude level was
computed  as the ratio of the number of recovered artificial stars to the 
number of added ones, and is given in  Table 1.  

The difference between input and output magnitudes of the artificial stars
provide an estimate of the photometric error alternative to the Daophot 
$\sigma$. The average
value associated to each magnitude bin is smaller than the corresponding
$\sigma$'s, as expected when errors do not come from blends due to
very crowded conditions, so we 
adopted the Daophot $\sigma$'s for the synthetic CMDs (see Section 4).

\begin{figure*}
\vspace{9cm}
\includegraphics{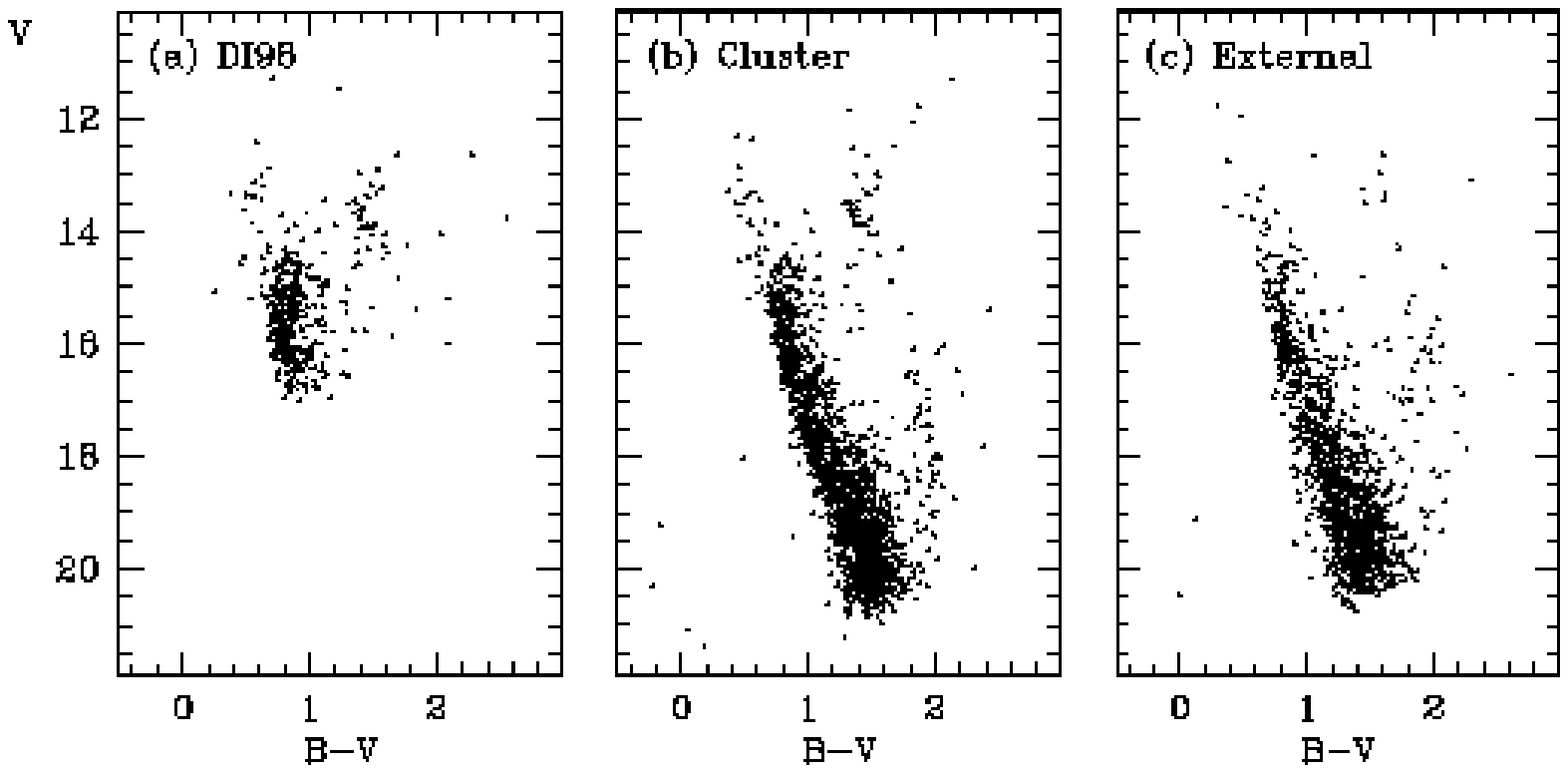}
\caption{(a) DI98 photographic CMD; (b) CMD of our central field; (c) 
CMD of our comparison field, which partly overlaps the central one}
\label{fig-cmd}
\end{figure*}

\section{The colour - magnitude diagram}

The final, calibrated sample consists of 1854 stars for which at least one
$B$ and one $V$ measures were obtained; as already said in the previous Section,
whenever more than one value was available
we adopted the average (after $\sigma$ clipping) of all measurements.
The table with $B$ and $V$ magnitudes and equatorial coordinates is available 
through the BDA (Mermilliod 1995: {\em obswww.unige.ch/webda/}).
The corresponding colour - magnitude diagram (CMD) is shown in 
Fig.~\ref{fig-cmd}(b), and can be compared with the DI98 one of
Fig.~\ref{fig-cmd}(a).

The Turn-Off point from the main sequence (MSTO) is located at 
$V \simeq 15.25$, and $B-V \simeq 0.64$,
and a red clump (the core He burning phase) is well visible, at 
$V \simeq 13.6$ and $B-V \simeq 1.33$.
That this second feature is attributable to the cluster appears clear when
looking at the external comparison field (Fig.~\ref{fig-cmd}(c)) where no clump
stars (or very few, as expected since there is some overlap) are visible.

\begin{figure*}
\vspace{15cm}
\includegraphics{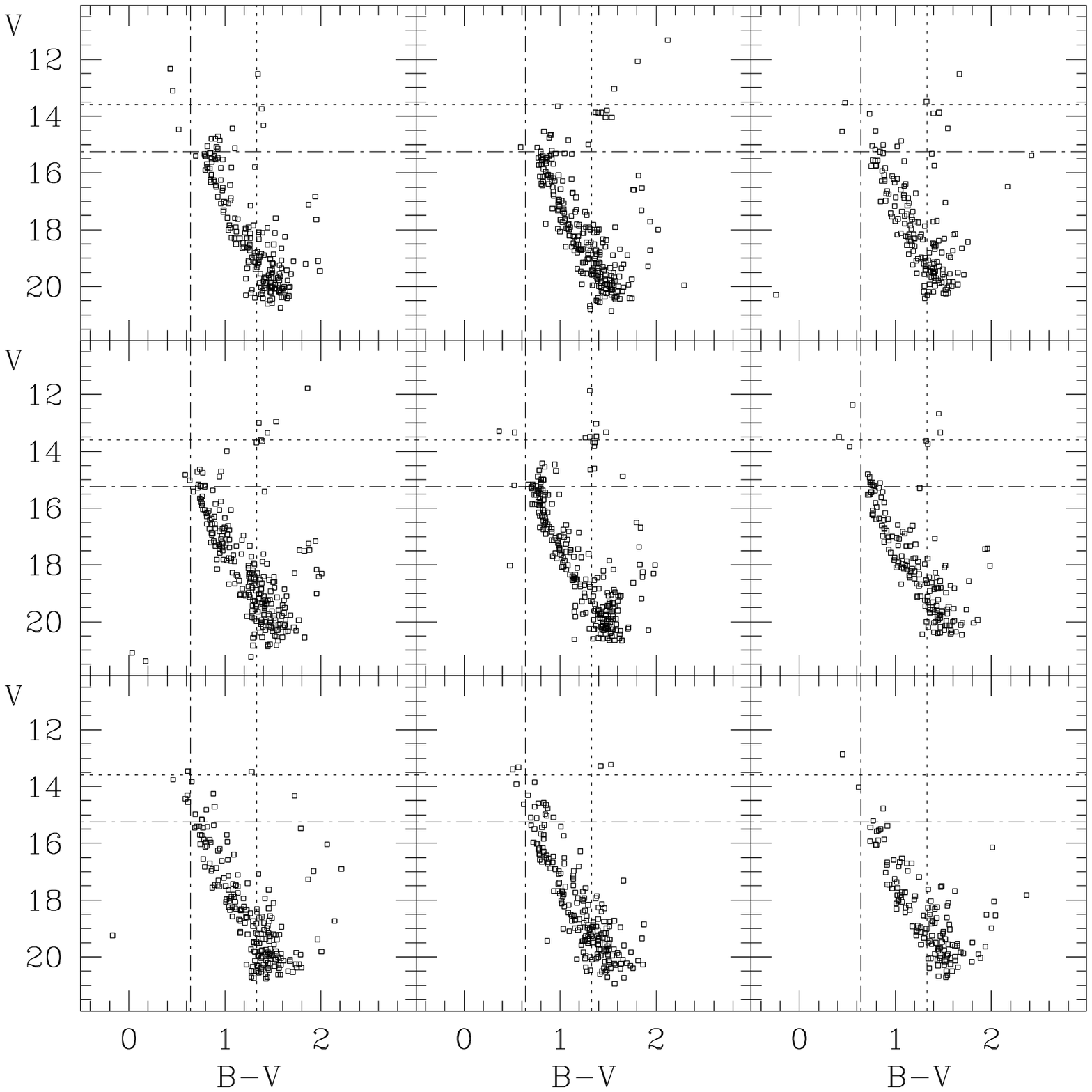}
\caption{The central field has been divided in a grid of nine 
700 $\times$ 700 pixels zones, and a CMD is plotted for each of them. The
lines are drawn to help identify the TO and clump positions.
Notice how the field contamination is negligible only in the
very central panel.}
\label{fig-var}
\end{figure*}

The cluster MS is well defined, but embedded in a fairly large strip due to
the fact that the
cluster is strongly contaminated by field stars. Since we also have
a second pointing not centered on the cluster, we could try to statistically
subtract field stars and isolate the cluster sequences. This was done, but
results were not satisfactory, given the small numbers involved and the fact
that the second field is still too close to the cluster centre and probably
contains some members. We then tried
another approach, and divided the central field in nine boxes of about 
700 $\times$ 700 pixels, plotting
separately the CMD's for each zone (Fig.~\ref{fig-var}). As it can be seen, the
cluster main sequence stands clearly over the field component only in the very
central zone: we decided to consider only stars falling
in this reduced field as {\it bona fide} cluster members, to be compared  with 
theoretical synthetic populations. 

DI98 also determined proper motions for about 60 \% of stars with $V, B-V$
data (their tab.5), but no clear separation between cluster and field stars
is visible from their data (their fig.2). They label as field stars those 
with proper motions in either coordinate larger than 1.2 arcsec per century,
but this criterium is not stringent enough: it excludes only about 10 \% of 
the sample, without any improvement in the CMD. We then decided to disregard
the proper motion information in the following.

\section{Cluster parameters}

In order to derive the age, distance and reddening of Cr~110, we have applied 
the synthetic CMD method (Tosi el al. 1991) already used and described for the 
other clusters of our sample. We have adopted as reference observational CMD
that of the central box of Fig.~\ref{fig-var}, which contains 260 stars. The
synthetic CMDs therefore contain 260 stars extracted with a MonteCarlo
procedure from the adopted sets of stellar evolution tracks. The synthetic
stars are affected by the
same photometric errors as the actual data and selected according to the
completeness factors listed in Section 2.2 (Table 1). The transformations
from the theoretical luminosity and effective temperature to the Johnson
magnitudes and colours have been performed  using Bessel, Castelli \& Pletz
(1998) conversion tables.

We have amply demonstrated in the past that the derived parameter values
depend on the adopted stellar evolution models. Hence, to give an
estimate of the corresponding uncertainties, one should always derive them
assuming various sets of stellar models computed with different assumptions.
For Cr~110 we have adopted homogeneous sets of  stellar tracks
computed  by three different groups: i) the Padova
models (hereinafter BBC, Bressan et al. 1993, Fagotto et al. 1994), which
take into account the effect of overshooting from convective regions, ii)
the FST tracks (Ventura, D'Antona \& Mazzitelli 2003 in preparation), with
various amounts of overshooting, and iii) the FRANEC tracks (hereinafter FRA,
Dominguez et al. 1999), computed without overshooting. Since the metallicity
of Cr~110 is unknown, for all these sets we have considered both solar and
sub-solar initial compositions. In practice we have simulated the observed
CMDs adopting the stellar models listed in Table 2. 

\begin{table}
\begin{center}
\caption{Stellar evolution models adopted for the synthetic CMDs}
\vspace{5mm}
\begin{tabular}{cccl}
\hline\hline
   Set  &metallicity & overshooting & Reference \\
\hline
BBC & 0.02 & yes &Bressan et al. 1993 \\
BBC & 0.008& yes &Fagotto et al. 1994 \\
BBC & 0.004& yes &Fagotto et al. 1994 \\
FRA & 0.02 & no &Dominguez et al. 1999 \\
FRA & 0.01 & no &Dominguez et al. 1999 \\
FRA & 0.006 & no &Dominguez et al. 1999 \\
FST & 0.02 & $\eta$=0.02 &Ventura et al. in prep.\\
FST & 0.02 & $\eta$=0.03 &Ventura et al. in prep.\\
FST & 0.006 & $\eta$=0.02 &Ventura et al. in prep.\\
FST & 0.006 & $\eta$=0.03 &Ventura et al. in prep.\\
\hline
\end{tabular}
\end{center}
\label{tracks}
\end{table}

All the synthetic CMDs have been computed either assuming that all the cluster
stars are single objects or that a fraction of them are members of binary 
systems. The data on Cr~110 don't allow to safely infer such fraction, 
but we find that one around 20\% seems to better reproduce the observed
distribution and spread of the cluster main sequence.
Indeed, despite the uncertainties, models with lower binary fractions lead
to excessively tight MSs, while models with 30\% begin to show a MS spread
larger than observed.
All the CMDs shown
in the figures assume that 20\% of the stars are binaries with random mass 
ratios.

\begin{figure*}
\vspace{9cm}
\includegraphics{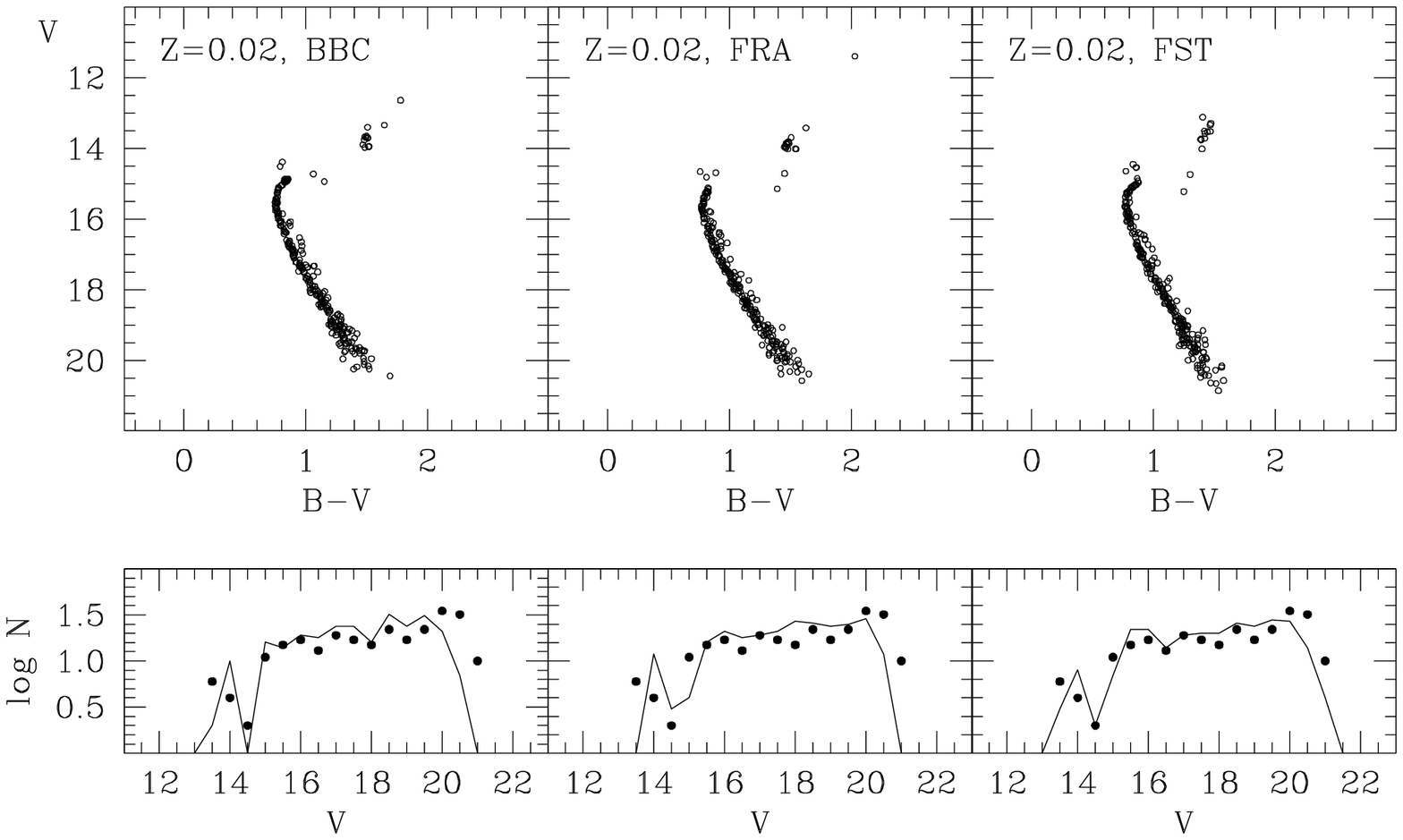}
\caption{Synthetic CMDs (top) and LFs (lines in the bottom panels) resulting 
from solar metallicity models. The dots in the bottom panels show the
empirical LF of the 260 stars in the CMD of the central panel of 
Fig.~\ref{fig-var}.
The displayed cases are: BBC models assuming age=1.35 Gyr, \ebv=0.40, \mmm=11.8,
on the left; FRA models assuming age=1.1 Gyr, \ebv=0.45, \mmm=11.9, on the 
center, and FST models with $\eta$=0.02 assuming age=1.5 Gyr, \ebv=0.38, 
\mmm=11.8, on the right.
}
\label{simsolar}
\end{figure*}

The relatively low number of stars measured in this cluster, especially in key
regions of the CMD such as the MSTO, the clump and the red giant branch (RGB),
in principle could not let us firmly discriminate between similar synthetic 
cases. Furthermore, the empirical luminosity function (LF) of the 260 supposed
cluster members may be somewhat contaminated by background objects,
especially at its faint end, which adds another uncertainty to the selection.
Nonetheless, the shape and the number of stars present in the  key 
evolutionary phases, as well as the magnitude and colour difference between
MSTO and clump, allow to derive a quite restricted range of possible ages for
each adopted set of tracks. Moreover, the stable results obtained for each 
assumed metallicity on distance modulus and reddening allow us to reach firm 
conclusions on these values too.  

Fig.~\ref{simsolar} shows the best cases of synthetic CMDs (top) and LFs
(bottom) resulting from solar metallicity tracks. 
The left hand panels refer to the case with BBC tracks, age=1.35 Gyr,
\ebv=0.40 and \mmm=11.8, the central ones to the case with FRA tracks,
age=1.1 Gyr, \ebv= 0.45 and \mmm=11.9, and the right hand panels to the
case with the FST tracks ($\eta$=0.02), age=1.5 Gyr, \ebv=0.38 and \mmm=11.8.
 
All the models with solar metallicity need to assume a reddening systematically
lower than the literature value (\ebv=0.5\footnote{Actually, DI98 derived
\ebv = 0.53 from the UBV two colours diagram for 22 probable cluster members,
while 0.50 is the value derived from isochrone fitting.}) to reproduce 
the observed colours. 
\ebv=0.45 is the maximum acceptable value, but 0.38--0.40 is more frequently 
found. As a consequence, the resulting distance modulus is systematically 
larger than that derived by DI98 (\mmm=11.45), and never shorter than 
\mmm=11.6.
Vice versa, models with sub-solar metallicity allow for larger reddenings
and for distance moduli in agreement with that  by DI98. 
 
\begin{figure*}
\vspace{9cm}
\includegraphics{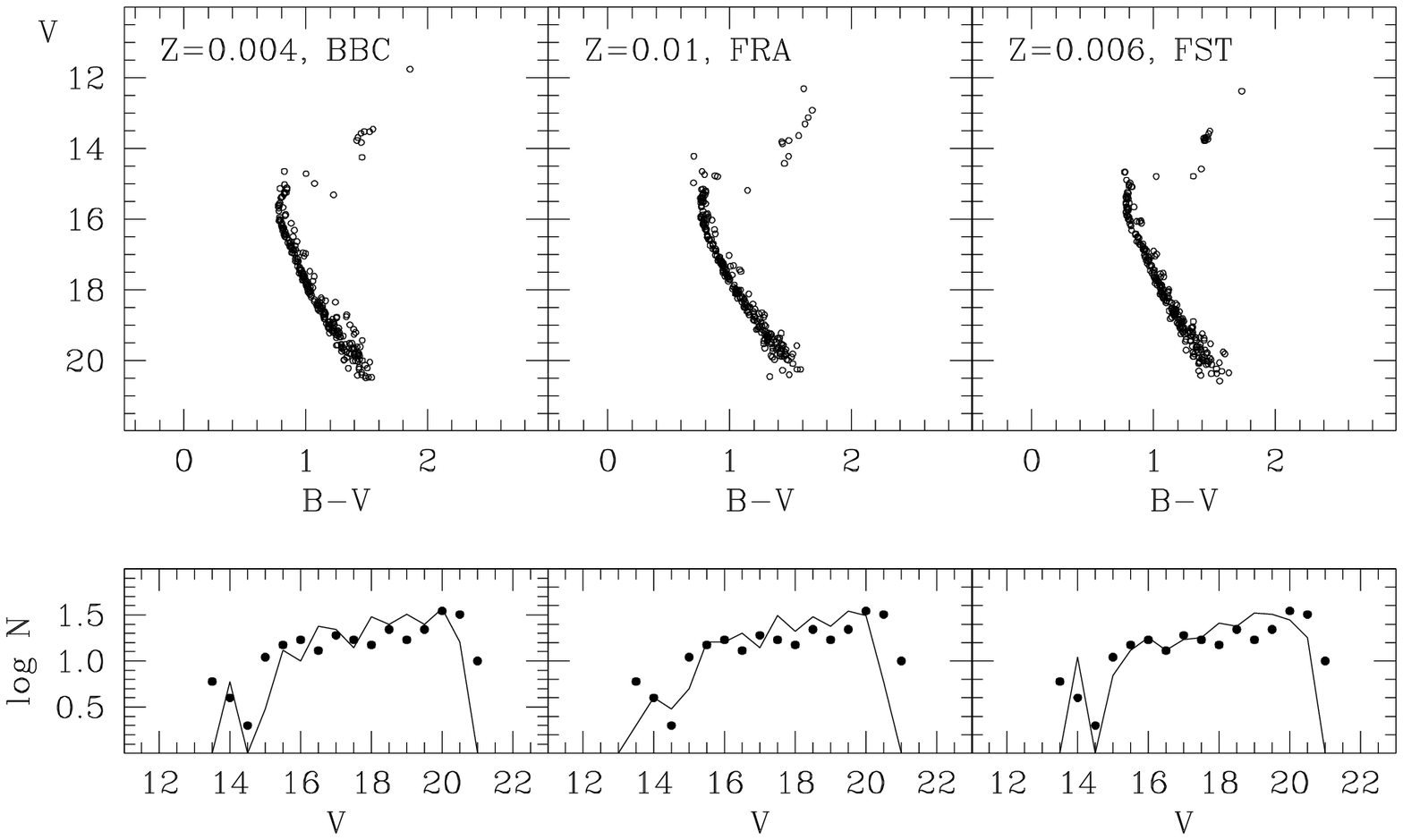}
\caption{Synthetic CMDs (top) and LFs (lines in the bottom panels) resulting 
from models with initial metallicity lower than solar. 
The dots in the bottom panels show the
empirical LF of the stars in the CMD of the central panel of Fig.~\ref{fig-var}.
The displayed cases are: BBC tracks with Z=0.004, age=1.7 Gyr, \ebv=0.57 and 
\mmm=11.45, on the left; FRA models assuming Z=0.01, age=1.2 Gyr, \ebv=0.52  
and \mmm=11.7, on the center, and FST models assuming $\eta$=0.02, Z=0.006, 
age=1.5 Gyr, \ebv=0.56 and \mmm=11.45., on the right.
}
\label{simsubsolar}
\end{figure*}

Fig.~\ref{simsubsolar} shows representative cases of synthetic CMDs (top) and 
LFs (bottom) resulting from stellar models with initial metallicity lower
than solar and in better agreement with the data. 
The left hand panels refer to the case with BBC tracks with
Z=0.004, age=1.7 Gyr, \ebv=0.57 and \mmm=11.45, the central ones to the case 
with FRA tracks with Z=0.01, age=1.2 Gyr, \ebv=0.52  and \mmm=11.7, and the 
right hand panels to the case with the FST tracks ($\eta$=0.02) with Z=0.006, 
age=1.5 Gyr, \ebv=0.56 and \mmm=11.45.

The main difference between the values we obtain with different stellar models
is the consequence of the different treatments of the convective zones.
Indeed, as well known, the age resulting from stellar models is increasingly 
older for increasing amount of assumed overshooting, due to the fact that 
larger cores make the model star brighter. In turn, for a given metallicity,
older ages correspond to predicted redder intrinsic MSTO colours, and therefore
need lower \ebv to reproduce the observed colour.

In general, we find that a better agreement with the observed cluster features
is obtained when assuming a metallicity lower than solar. The shape and the 
stellar distribution in the CMD of MS, MSTO, clump and RGB as well as
the LFs are all better reproduced by low metallicity models. How much
lower than solar is difficult to say, because the formal metallicity of
the stellar tracks is actually a combination of chemical composition,
opacities, etc., and therefore its effect may vary from one set to another.
In fact, we find that, when adopting the BBC models, Z=0.004 provides better
results than Z=0.008, while with the FRA models Z=0.01 looks better
than Z=0.006, but Z=0.006 is fine with the FST models. These apparent
inconsistencies are simply the proof that photometric use of stellar models
alone is not a good metallicity indicator; at most it can be useful to
infer a metallicity ranking. No doubt that spectroscopic observations of
Cr~110 are needed to safely derive its true chemical abundance
and restrict the choices for the other cluster parameters.

\section{Summary and conclusions}

Cr~110 is a fairly poor cluster, highly contaminated by background objects,
with the further disadvantage that the latter have a distribution in the CMD
not too different from that of the cluster members, a circumstance that makes
separation of true cluster members particularly uncertain. In spite of that, we
are able to attribute to  Cr~110 fairly restricted ranges of age,  reddening
and distance modulus, whose uncertainty is mostly related to the different
possible theoretical approaches and to the unknown metallicity. If the cluster
metallicity is solar, its age ranges between 1.1 Gyr (based on stellar models
with no overshooting) and 1.5 Gyr (based on models with a large amount of
overshooting), the reddening is \ebv around 0.4 and the distance modulus
larger than 11.6.
If the metallicity is less than solar, the age ranges between
1.2 Gyr (no overshooting) and 1.7 Gyr (overshooting), the reddening is
between 0.52 and 0.58 and the distance modulus 11.45 or slightly larger.

We suggest that the cluster metallicity be lower than solar because the
corresponding synthetic CMDs are more capable to reproduce the shape of
the various evolutionary sequences, their number of populating stars and
their relative colour and magnitude differences. In addition, with a
metal poor chemical composition we obtain values for the reddening and the
distance modulus in much better agreement with those derived by DI98.
The metal poor synthetic CMDs in better agreement with the data assume
in fact 0.52$\leq$ \ebv $\leq$0.57 and 11.45$\leq$ \mmm $\leq$11.7.

Better and more stringent constraints require further observations:
spectroscopy at (at least) medium resolution of a large sample of objects to
determine membership, and at high resolution of a few member stars to measure
detailed chemical abundances. This way Cr 110 could appropriately take its place
in our slowly growing sample of well studied open clusters. 

\bigskip\noindent
ACKNOWLEDGEMENTS

This work is based on observations collected at the European Southern
Observatory, Chile.
We warmly thank P. Montegriffo, whose programs were used for the data analysis,
and E. Sabbi for her expert advice.
We also thank L. Di Fabrizio and G. Clementini for providing the standard stars
photometry and the extinction coefficients, and F. D'Antona and P. Ventura for
providing their unpublished stellar models.
The bulk of the simulation code was originally provided by L.Greggio.
Financial support to this project has come from the MURST-MIUR through 
Cofin98 ''Stellar Evolution'', and Cofin00 ''Stellar Observables of Cosmological
Relevance''.
This research has made use of the Simbad database, operated at CDS, Strasbourg,
France. 
Finally we acknowledge the use of the valuable BDA database, maintained by
J.-C. Mermilliod, Geneva.

\end{document}